\documentclass[twocolumn,pre,floats,aps,amsmath,amssymb]{revtex4}
\usepackage{graphicx}
\usepackage{bm}
\usepackage[section]{placeins}

\begin{document}

\title{Compact chromium oxide thin film resistors for use in nanoscale quantum circuits}
\author{C. R. Nash, J. C. Fenton, N. G. N. Constantino, P. A. Warburton}
\affiliation{London Centre for Nanotechnology, UCL, 17-19 Gordon Street, London, WC1H 0AH}
\date{\today}

\begin{abstract}

We report on the electrical characterisation of a series of thin chromium oxide films, grown by dc sputtering, to evaluate their suitability for use as on-chip resistors in nanoelectronics. By increasing the level of oxygen doping, the room-temperature sheet resistance of the chromium oxide films was varied from 28\,$\Omega / \square$ to 32.6\,k$\Omega / \square$. The variation in resistance with cooling to 4.2\,K in liquid helium was investigated; the sheet resistance at 4.2\,K varied with composition from 65\,$\Omega / \square$ to above 20\,G$\Omega / \square$. All of the films measured displayed ohmic behaviour at all measured temperatures. For on-chip devices for quantum phase-slip measurements using niobium-silicon nanowires, interfaces between niobium-silicon and chromium oxide are required. By characterising the interface contact resistance, we found that a gold intermediate layer is favourable: the specific contact resistivity of chromium-oxide-to-gold interfaces was 0.15\,m$\Omega$cm$^2$, much lower than the value for direct chromium-oxide to niobium-silicon interfaces, 65\,m$\Omega$cm$^2$. We conclude that these chromium oxide films are suitable for use in nanoscale circuits as high-value resistors, with resistivity tunable by oxygen content.
\end{abstract}

\maketitle

\section{Introduction}
\label{sec:intro}

Since the 1960s, on-chip thin-film resistors have found applications in integrated circuits \cite{Chapman, Hall}. Materials such as NiCr \cite{Bloch, Lai}, SiCr \cite{Waits, Jhabvala}, TaN \cite{Akashi, Malmros}, and TiO \cite{Lotkhov} have all been used, fabricated by evaporation or sputter deposition. They provide room-temperature sheet resistances ranging from 10\,$\Omega / \square$ to 2000\,$\Omega / \square$, but materials to provide higher sheet resistances are less commonly used, with W and Bi \cite{Lehtinen} being reported for this use. In nanoscale electronic circuits, there is frequently a need for resistive elements which are smaller than the wavelength associated with the frequency of interest. In particular, in coherent quantum circuits there is a need for compact resistors to isolate devices from the environment at frequencies in the GHz range \cite{Webster}. The isolating resistor must also be sufficiently compact that its resistance is not shorted by the capacitance across the substrate. Such resistors have found applications in the fields of Josephson junctions and nanoscale circuits \cite{Martinis, KuzminAug, KuzminNov, Zorin, Krupenin}. In order to provide a high-value resistor using standard materials, it is necessary to pattern long (often meandering) paths, thereby leading to an undesirably high shunt capacitance. 

A particular example of an application requiring high-value resistors is circuits for exploiting quantum phase-slips. In 2006, Mooij and Nazarov \cite{Mooij} showed a duality between Josephson junctions and a coherent quantum phase-slip (QPS) circuit element --- a superconducting nanowire --- which implies the potential for a new quantum standard for current. On-chip resistors play an important role in the current-standard QPS circuit, providing a shunt resistance to ensure the overdamped behaviour in which microwave-induced step features are best observed. Work on investigating quantum phase-slips in superconducting nanowires has ensued, including recent microwave spectroscopy measurements, which showed features of coherent QPS origin \cite{Astafiev}. The QPS current-standard circuit described by Mooij and Nazarov comprises two resistors on either side of a niobium-silicon nanowire, with a combined series resistance which should exceed a certain value in order to minimise hysteresis in the current-voltage characteristics of this particular circuit. For typical parameters, the series resistance should exceed 60\,k$\Omega$.

Chromium oxide is a potential candidate for use as a high-resistance thin-film component \cite{Zorin}. Chromium oxide has a resistance that is tunable through oxygen doping, as well as showing good adhesion to silicon substrates; it is also easy to pattern by chemical etching or lift-off.  The electrical and mechanical characteristics of chromium \cite{Gould, Tompkins, El-Hiti}, Cr$_2$O$_3$ \cite{Lewis}, and CrO$_2$ \cite{Ku} are all well-documented, due to the numerous applications of these materials in microcoating and magnetic tape manufacture. Less work has focused on amorphous chromium oxide films. Recent studies \cite{Pang2011, Pang2008, Lou} have looked into the mechanical properties of amorphous films and found good hardness and resistance to wear, but the electrical properties of the films are less well-documented. The films are non-magnetic and therefore suitable for use in a circuit with superconducting elements in close proximity. In the past, thin, weakly oxidised Cr resistors with a sheet resistance up to 800\,$\Omega / \square$ have been employed successfully to provide high resistance \cite{Zorin}; however, the small widths and large length scales necessary to create higher-value resistors in materials such as weakly oxidised chromium or NiCr can be more difficult to fabricate, in addition to the issue of parasitic capacitances. Strongly oxidised chromium oxide films have the advantage of a larger sheet resistance, meaning a high-resistance structure can be more compact. In this paper, the electrical properties of chromium oxide films are investigated in detail. We report on the fabrication of thin-film chromium oxide resistors by sputter deposition, and investigate their structural and electrical properties. 

It is also important to control the interfaces which thin-film resistors make to other materials. When two materials meet, there will usually be imperfect contact between them. Impurities, unevenness in each surface and the mismatch of lattice structures will cause voids and defects at the surface of the two materials, which will lead to an electrical interface resistance \cite{Chang}. Minimising this contact resistance is one of the main practical challenges in fabricating multi-layered on-chip circuits. We have therefore investigated interfaces between chromium oxide and other materials. We report on tests of niobium-silicon-to-chromium-oxide interfaces; these are relevant in the preparation of these resistors for use in a quantum phase-slip circuit in which the thin-film resistors must be electrically connected to niobium-silicon nanowires fabricated by multi-stage electron-beam lithography. We also present results of tests of chromium oxide-to-gold interfaces. 

\section{Composition}
\label{sec:experiment}

\subsection{Fabrication}

A series of chromium oxide films was created, using a broad range of oxygen dopant levels. The chromium oxide films were deposited on p-doped silicon by dc-magnetron sputtering in an argon and oxygen atmosphere, using a 3"~Cr target with a 500\,W sputter power for ten minutes with a three-minute pre-sputter. The target-to-substrate distance was 18\,cm. The argon gas pressure was 5\,mTorr. The film thicknesses, measured by a DektakXT surface profiler, were all in the range 179\,nm for growth in pure argon to 246\,nm for growth with an oxygen partial pressure of 0.7\,mTorr, implying deposition rates from 17.9\,nm/min to 24.6\,nm/min. 

\subsection{Structure}

\begin{figure}[htbp]
  \centering
    \includegraphics[width=0.5\textwidth]{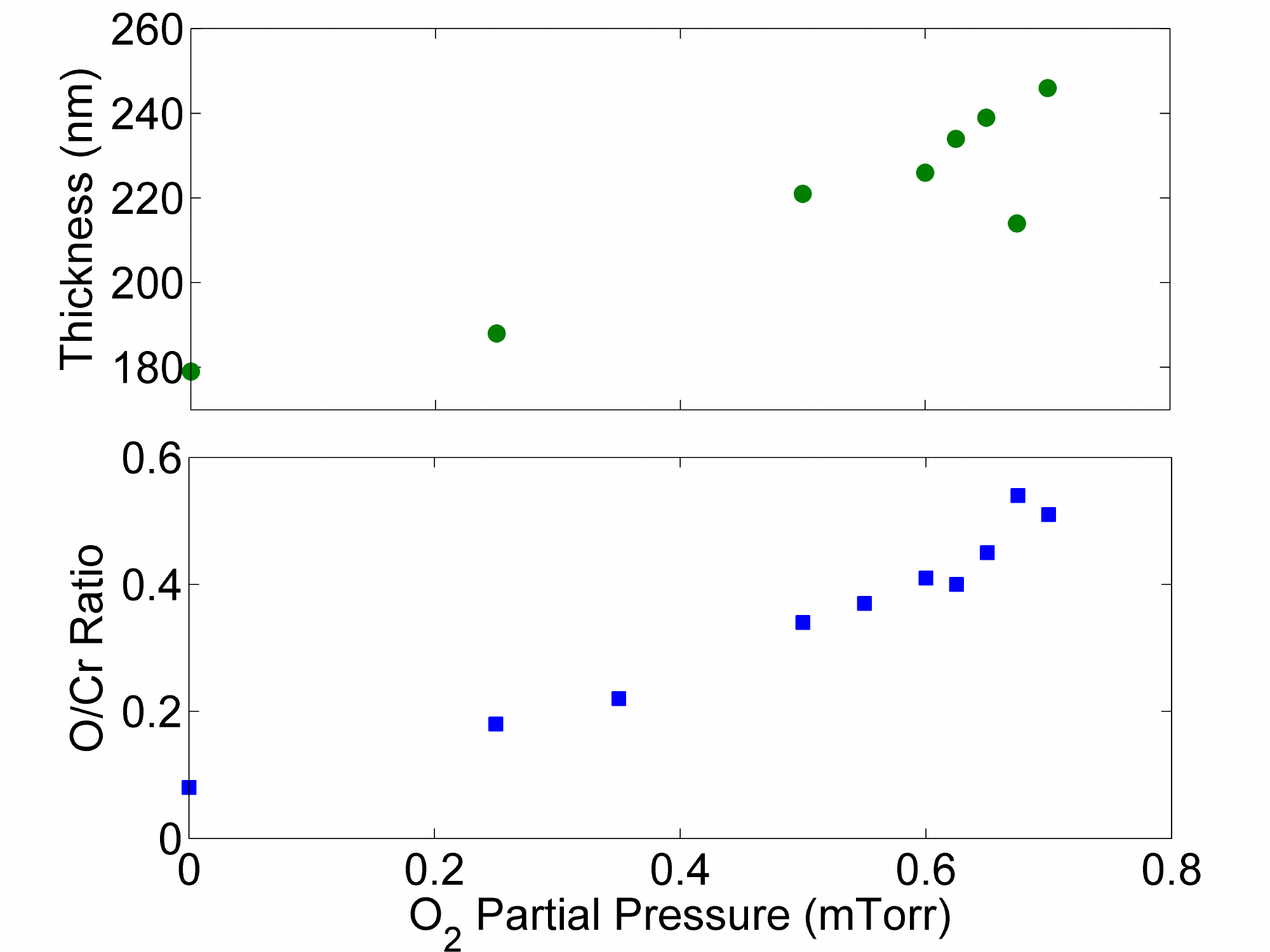}
  \caption{(Upper) Variation of the thickness of material, after a ten minute sputter time, with the oxygen partial pressure. (Lower) Dependence of the oxygen-to-chromium mass ratio in the film (as measured by WDS) on the O$_2$ partial pressure during growth. An argon gas partial pressure of 5\,mTorr was used for all samples.}
 \label{fig:OCrPressure}
\end{figure}

Wavelength-dispersive spectroscopy (WDS) in a scanning electron microscope was used to determine the composition of the films grown. Fig.\,\ref{fig:OCrPressure} shows the variation of the oxygen-to-chromium mass ratio in the film with the O$_2$ partial pressure used during sputter deposition. There is a gradual increase in the proportion of oxygen in the film as the pressure of oxygen introduced into the argon gas in the sputterer is increased. There is oxygen present in the film even if no oxygen is added whilst sputtering. This is likely due to oxidation of the surface of the film after removal from the chamber or the influence of residual oxygen in the chamber during growth. 

X-ray diffraction was used to investigate the structure of the films. All of the films from Fig.\,\ref{fig:OCrPressure} were measured and all were found to be amorphous, with only substrate peaks visible in all measurements. This shows that the magnetic phases of chromium oxide, which could cause suppression of superconductivity in adjacent nanowires, were not present in these films.

\section{Resistance measurements}
\label{sec:results}

\subsection{Variation in room-temperature resistance with oxygen partial pressure}

Four-point electrical transport measurements were performed at room temperature to determine the sheet resistance of the films. A linear contact configuration was used and the measurement was made immediately after removing the sample from the sputterer vacuum chamber.

\begin{figure}[htbp]
  \centering
    \includegraphics[width=0.5\textwidth]{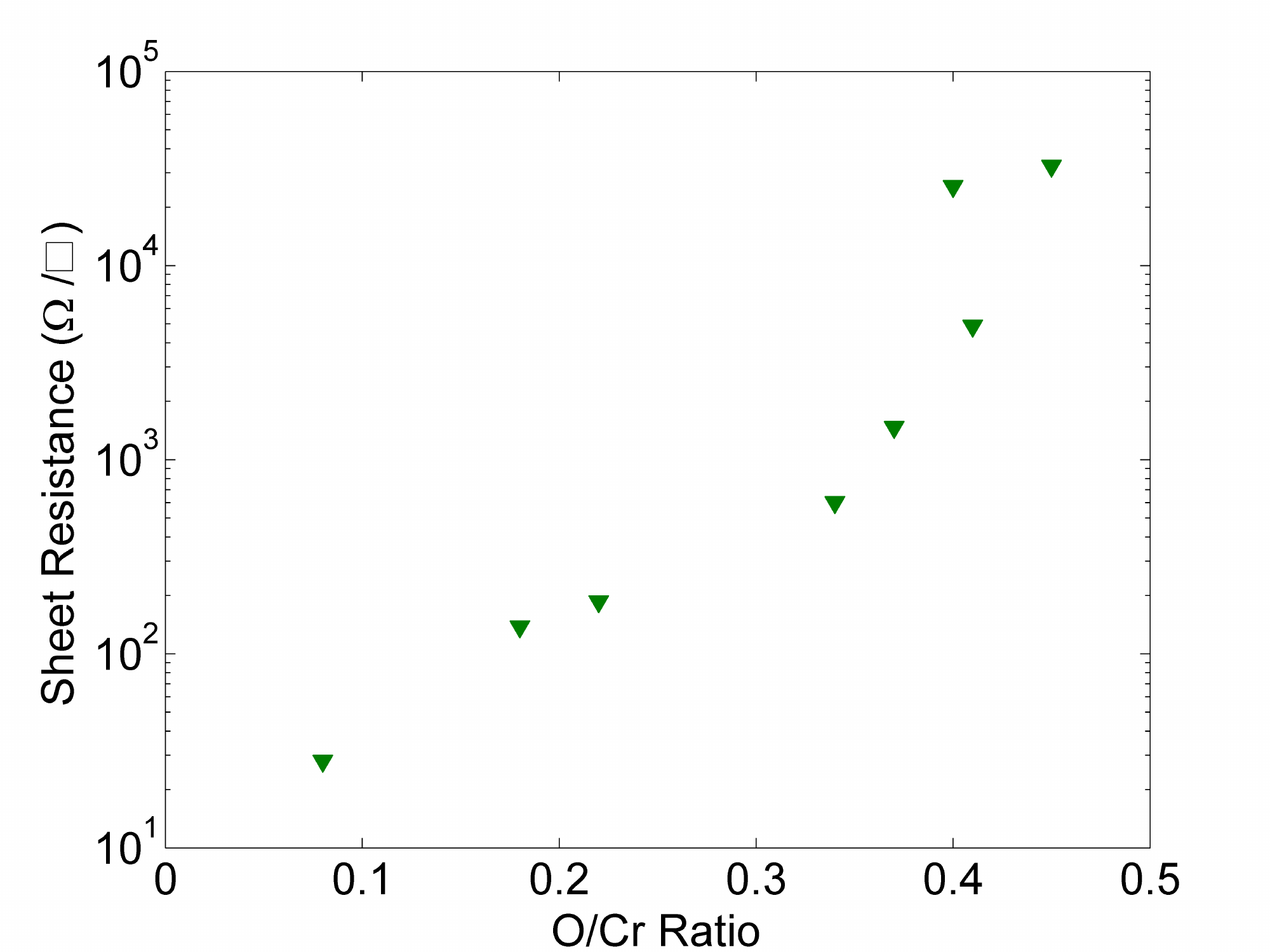}
  \caption{Variation of room-temperature sheet resistance of chromium oxide films with the oxygen-to-chromium mass ratio for the same samples as shown in Fig.\,\ref{fig:OCrPressure}.}
 \label{fig:SheetROCr}
\end{figure}

Fig.\,\ref{fig:SheetROCr} shows the variation of room-temperature sheet resistance of the chromium oxide films (the same samples as shown in Fig.\,\ref{fig:OCrPressure}) with the oxygen-to-chromium mass ratio \cite{Footnote}. There is a steep increase in room-temperature sheet resistance as the oxygen incorporation increases, corresponding to a departure from standard metallic conduction. The samples with O/Cr mass ratios of greater than 0.45 are not shown, as their resistances were higher than the measurement range (100\,k$\Omega$) of the particular equipment employed.

\subsection{Variation of resistance with temperature}

In order to measure the resistance of the films as a function of temperature, bonded aluminium wires were used to make contact to the films at four points, spaced by approximately 1\,mm, in a linear configuration and the resistance was measured by means of a four-point technique. At room temperature, conduction was ohmic in the measured range of currents, up to 200\,$\mu$A. The samples were cooled to 4.2\,K in liquid helium, and the variation of the resistance of the films with temperature during cooling was measured, using a dc bias current of 10 $\mu$A for all films. Conduction remained ohmic at 4.2K for all measured films. 

\begin{figure}[htbp]
  \centering
    \includegraphics[width=0.5\textwidth]{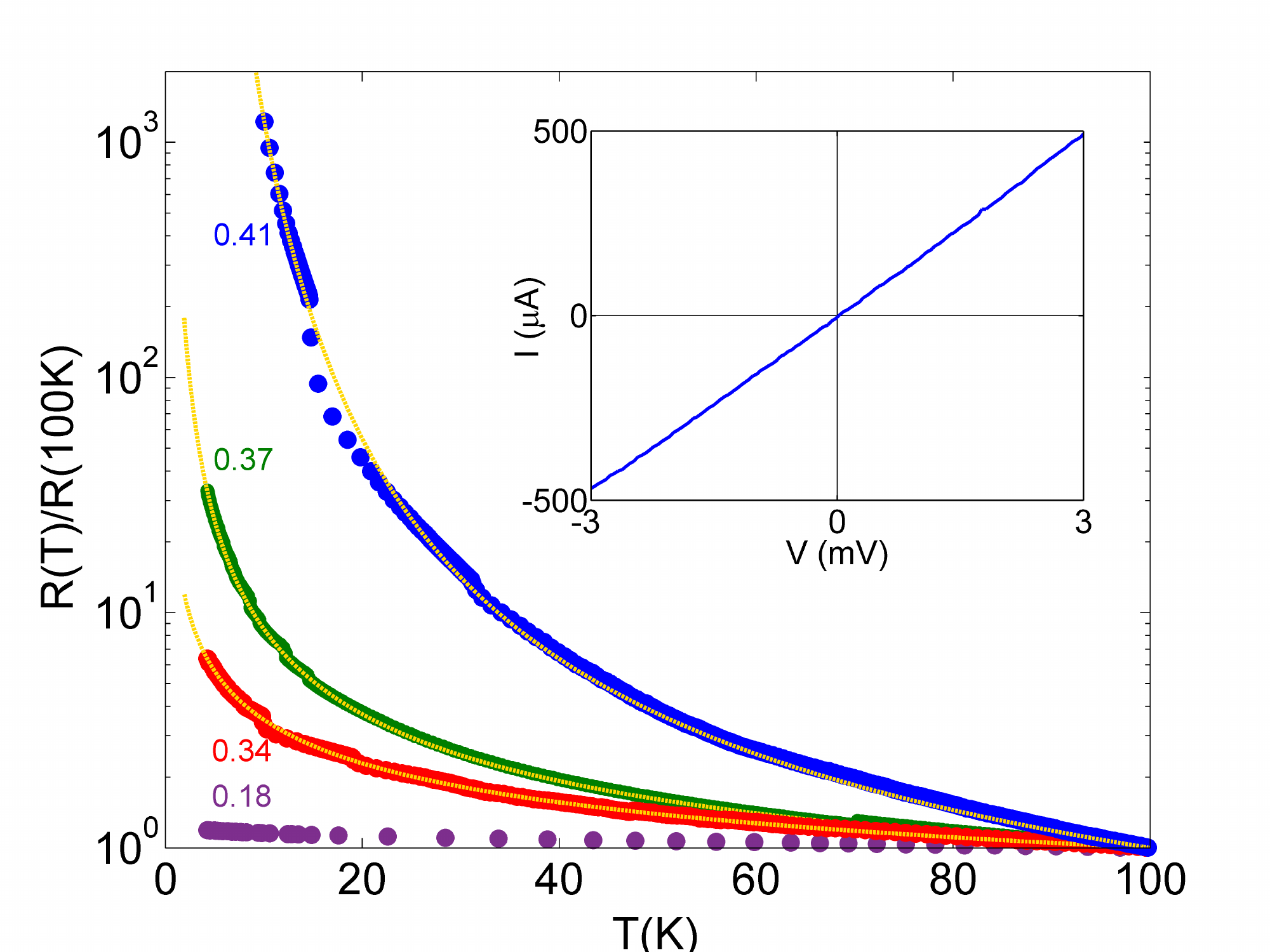}
  \caption{Variation with temperature of measured four-point resistance, normalised to the value at 100\,K, for chromium oxide films of varying composition. Labels denote the oxygen-to-chromium mass ratio of the film, as determined by WDS. Lines in gold denote fits to a variable-range hopping model. Films were biased at 10\,$\mu$A and cooled to 4.2\,K. Small jumps in the data are an artefact associated with thermal lag during the measurement. (Inset): IV curve of 0.41 O/Cr mass ratio film at 4.2\,K demonstrating ohmic behaviour.}
 \label{fig:RT}
\end{figure}

Fig.\,\ref{fig:RT} shows the measured variation of four-point resistance with temperature for chromium oxide films of varying composition. The data has been normalised with respect to the resistance at 100\,K; in the temperature range shown, conduction through the substrate is negligible. The graph shows that oxygen incorporation has a significant effect on the variation of resistance of the films with temperature, and there is a large variation between the films in the value of low-temperature resistance. The films become more resistive for increasing oxygen concentration. The sheet resistance at 4.2\,K was obtained from the measured resistance by means of a standard transformation for the films \cite{Smits} and is shown in Table 1. The 0.41 mass ratio film reaches a maximum sheet resistance greater than 20\,G$\Omega$ at 4.2\,K. More precise measurement of the resistance and IV characteristics at low temperature were not possible for this film because its resistance approached the input impedance of the measurement equipment. 

The data  shown in Fig.\,\ref{fig:RT} have been fitted to a variable-range hopping form, $R = R_{0}\exp(T_{0}/T)^{n}$, where $T_{0}$ is the localisation temperature. For Mott variable-range hopping (VRH), $n = (d+1)$ in $d$ dimensions; for Efros--Shklovskii VRH, appropriate for stronger electron-electron interactions, $n=1/2$ \cite{Zhang}. This functional form provides a good fit to the data for the films with higher oxygen content, as shown in Table 1. The fitted value of $n$ varies with the oxygen concentration; it is consistent with 3-D Mott VRH for the 0.37 O/Cr film and Efros--Shklovskii VRH for the 0.41 O/Cr film. The crossover from Mott VRH to Efros--Shklovskii VRH with increasing oxygen content is in line with expectations that electron-electron interactions become increasingly important in the most resistive films. The values of $T_0$ were found to be of the order of those found in other materials that exhibit Efros--Shklovskii VRH \cite{Abraham, Essaleh}.  $T_{0}$ is related to the localisation length $\xi$ via the equation $T_{0} =e^2\beta/k_Bk\xi$ where $\beta$ is a numerical constant, $k_B$ is the Boltzman constant and $k = \epsilon_r\epsilon_0$ is the electrical permittivity of the material \cite{Mott, Hill}. Dielectric constants for disordered metals have not generally been reliably determined, only estimated \cite{Abraham, Butko}; independent of such estimates $\epsilon_r\xi$ can be calculated from $T_{0}$. Taking $\beta = 2.8$ \cite{Butko}, we obtain $\epsilon_r\xi =1240$\,nm for the 0.37 mass-ratio film and $\epsilon_r\xi =880$\,nm for the 0.41 mass-ratio film. This is consistent with previous work \cite{Abraham} where $\xi$ was found to be in the range of tens to hundreds of angstroms when $\epsilon_r$ was estimated as $10^4$. The high value of $T_{0}$ for the 0.34 mass-ratio film may indicate that this film is at the border of applicability of this model. Poor fits for the lower oxygen-content films indicate that conduction in those films is too metallic to be accurately predicted by the VRH model.

\begin{table}[htbp]
\centering
\begin{tabular}{l*{3}{c}r}
O/Cr            & & & & \\
Mass           & $R_{(4.2\,K)}$  & $n$ & $T_0$ &  Fit  \\
Ratio 	  & ($\Omega / \square$) & & (K) & $\mathcal{R}^2$ \\
\hline
0.18		  & 65 & - & - &  - \\
0.22             & 317 & - & - &  - \\
0.34             & 1.1$\times$10$^4$ & 0.185 ($\pm$\,0.006) & 41490 &  0.9996 \\
0.37             & 9.5$\times$10$^4$ & 0.351 ($\pm$\,0.004) & 474 &  0.9998  \\
0.41             & $>$2.0$\times$10$^{10}$ & 0.534 ($\pm$\,0.014) &  818 & 0.9988  \\
\end{tabular}
\caption{Variation of the measured four-point resistance at 4.2\,K for chromium oxide films of varying oxygen concentrations. $R_{(4.2\,K)}$ is the measured sheet resistance at 4.2\,K. Coefficients of a fit to the equation $\ln{R} = \ln{R_{0}} + (T_{0}/T)^{n}$ are also shown. $\mathcal{R}^2$ is the coefficient of determination and indicates the goodness of fit. Fits for the 0.18 and 0.22 mass-ratio films to this model were poor, associated with little change in resistance across the temperature range.}
\end{table}
 
\section{Resistance of chromium oxide interfaces for the development of a QPS circuit}
\label{sec:contact}

When connecting chromium oxide resistors to niobium-silicon nanowires in a QPS circuit, there are several possible methods to use. Direct connection of the components may be used by overlapping the components during deposition; this has the advantage of minimising the number of fabrication steps necessary. An alternative is to use an inert non-oxidising interlayer, such as gold, to connect the two. This might be appropriate if direct contact would result in too high a contact resistance.

\begin{figure}
  \centering
    \includegraphics[width=0.45\textwidth]{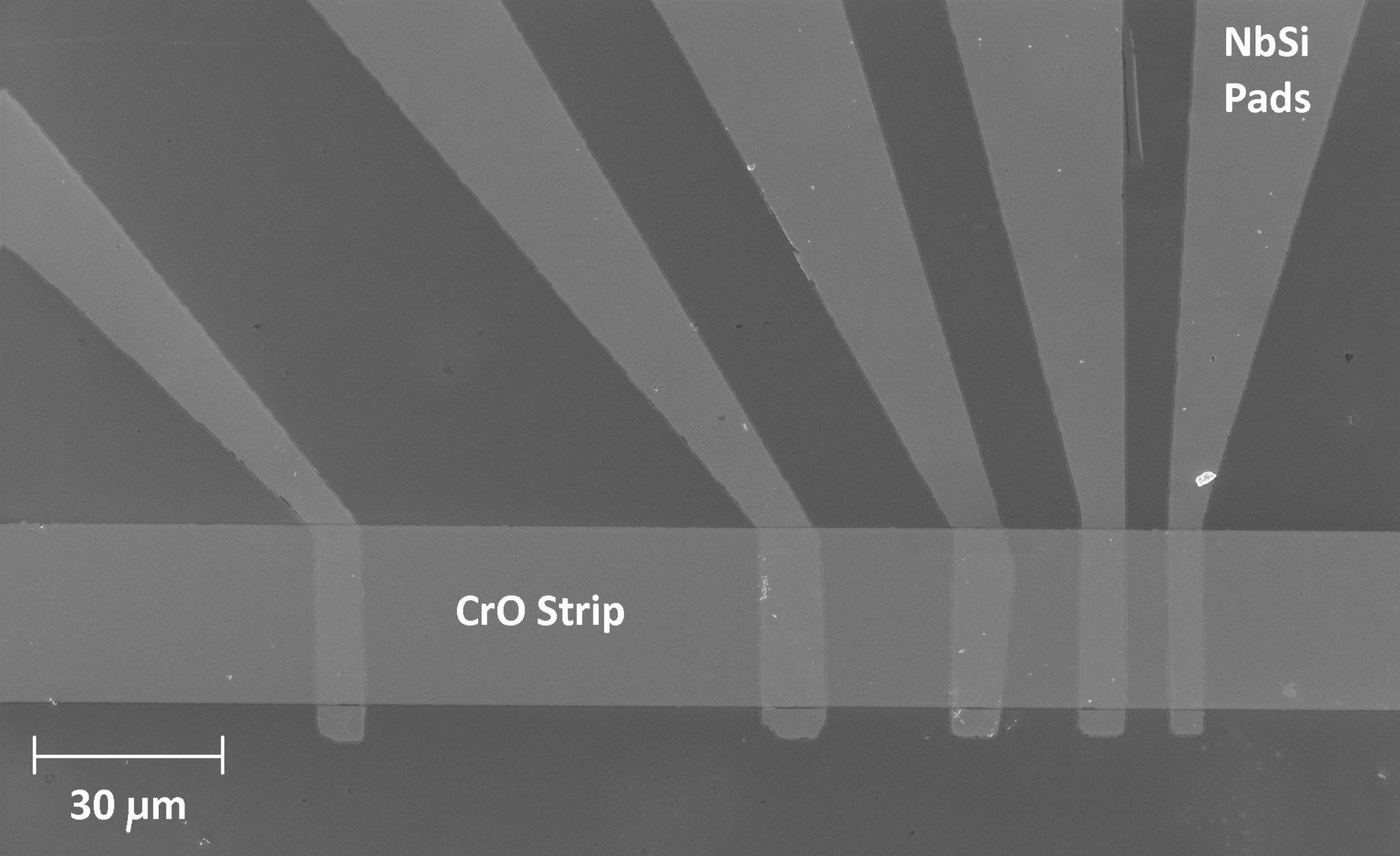}
  \caption{Optical image of a contact-resistance test pattern fabricated with photolithography. Niobium-silicon pads of width 5\,$\mu$m ($\pm$1\,$\mu$m) are in contact with a chromium oxide strip of width 30\,$\mu$m.}
 \label{fig:Contact}
\end{figure}

In order to determine whether QPS nanowires could be connected in series to chromium oxide resistors without a large contact resistance, the contact resistance of the interface between chromium oxide and niobium-silicon was investigated. A transmission-line model test pattern was used to determine the contact resistance \cite{Reeves}. This model pattern consists of a series of pads of one material spaced apart by varying distances, overlaid by a strip of another material (Fig.\,\ref{fig:Contact}). The resistance between any two of these pads is  $R \simeq 2R_{contact} + R_{strip}$, where $R_{contact}$ is the contact resistance between the pad and the strip, $R_{strip} = r_{strip}\,D$ is the resistance of the strip material,  $r$$_{strip}$ is the resistance per unit length of the strip material and $D$ is the length of strip between the two pads. Two-point measurements are performed for each pad combination and the known resistance of the pads is subtracted from this value in order to determine $R$. On a plot of $R$ against $D$, the $D = 0$ intercept conveniently gives the contact resistance. 

The transmission-line model test patterns were fabricated using photolithography. Initially, niobium-silicon was used as the pad material and chromium oxide was used as the strip material. First, a pattern of niobium-silicon pads was created by lithography. Niobium-silicon pads were co-deposited by sputtering using a 2"~Nb target and a 2"~Si target, both with a target-to-substrate distance of 15\,cm. Silicon was sputtered at 150\,W while niobium was simultaneously deposited at 75\,W, with a sputter time of five minutes, leading to a film with a composition of Nb$_{0.2}$Si$_{0.8}$ and a thickness of 70\,nm.  

After lift-off, a second lift-off mask and chromium oxide sputter deposition were used to produce a chromium oxide strip with a mass ratio of around 0.37\,O/Cr. The area of each niobium-silicon-to-chromium-oxide contact was 5$\times$30\,$\mu$m$^2$. Aluminium bond wires were connected from the sample pads to copper pads that led to the measurement equipment. Two-terminal resistance measurements were made in the range $\pm$10$\mu$A between every pad combination at 4.2\,K. The measured resistance is dominated by the contact resistance and resistance of the strip.  

\begin{figure}
  \centering
    \includegraphics[width=0.5\textwidth]{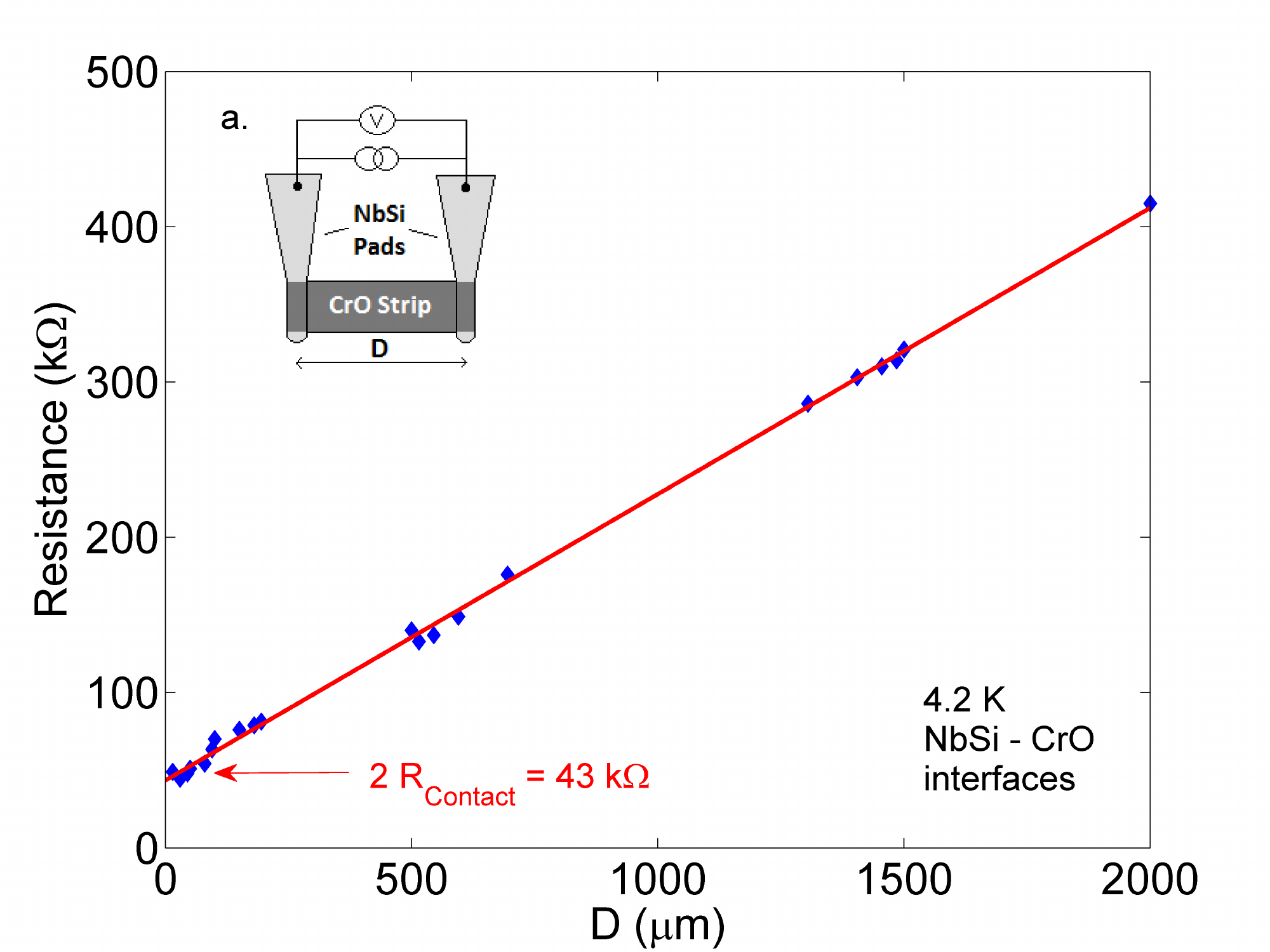}
  
	\includegraphics[width=0.5\textwidth]{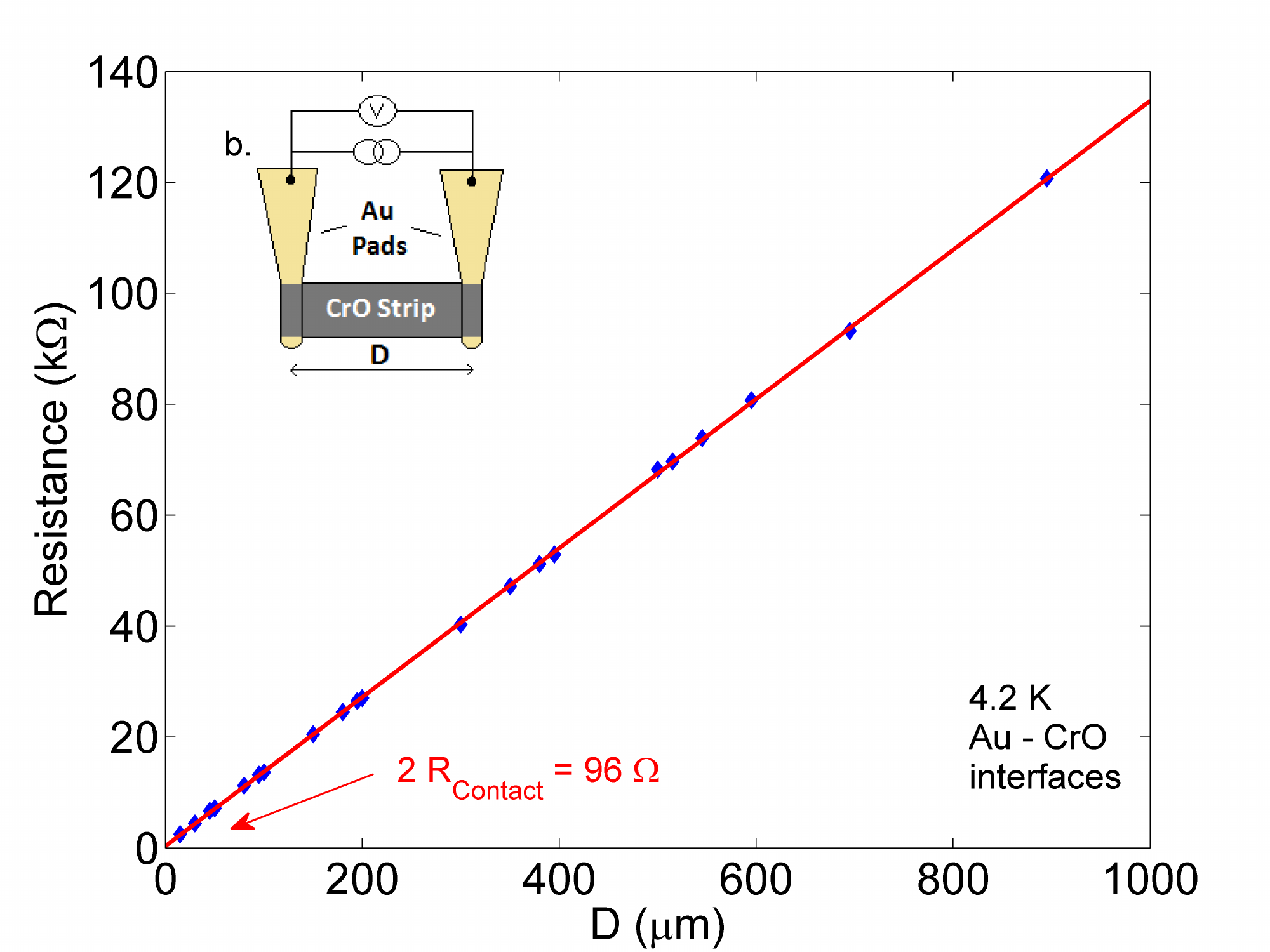}
	\caption{(a). Variation of two-terminal contact resistance at 4.2\,K with distance between contacts for a contact test pattern of NbSi (pads) and chromium oxide (strip). (b). Variation of two-terminal contact resistance with distance between contacts for a contact test pattern of gold (pads) and chromium oxide (strip). (Inset): Experimental set-up for the two-terminal measurements of each pair of pads. The resistance of the pads has been subtracted from the data before plotting.}

 \label{fig:NbSiContact}
\end{figure}

Fig.\,\ref{fig:NbSiContact}(a) shows the variation of the two-terminal resistance as a function of the distance between the pads for the niobium-silicon-to-chromium-oxide interfaces. $R_{contact}$ is 22\,k$\Omega$, implying a large specific contact resistivity of 65\,m$\Omega$cm$^2$ (6.5\,M$\Omega\mu$m$^2$). For use in a circuit for QPS current-standard experiments a contact area of a few square microns is generally required. The contact resistance for this interface would exceed the chromium oxide resistance, and so is undesirably high.

Utilising a gold interlayer is a potential solution to this problem; we therefore investigated the contact resistance between chromium oxide and gold. To produce gold contact pads, a 10\,nm chromium adhesion layer, and afterwards a 40\,nm gold layer, were thermally evaporated and then patterned using a positive photolithography resist to define the pad shape. A wet-etch method was then used to remove the unwanted metal and a solvent used to remove any remaining resist. The chromium oxide strip was then added, as before, using sputtering and a lift-off process. Fig.\,\ref{fig:NbSiContact}\,(b) shows that the contact resistance of gold to chromium oxide was 48\,$\Omega$, implying a specific contact resistivity of 0.15\,m$\Omega$cm$^2$ (15\,k$\Omega\mu$m$^2$). This is several orders of magnitude less than the contact resistance between niobium-silicon and chromium oxide. By using gold as a intermediate layer, the contact resistance at interfaces between different materials can therefore be minimised. 

\section{Conclusion}
\label{sec:conclusion}

We have demonstrated that it is possible to reliably control the resistance of chromium oxide films at cryogenic temperatures by controlling the oxygen pressure at the time of deposition. Films can be sputtered with low-temperature sheet resistances in the range of M$\Omega$--G$\Omega$, far higher than currently obtained with standard thin-film resistor materials. For example two resistors with an O/Cr mass ratio of 0.34, thickness 100\,nm, width of 65\,nm and length of 2.1\,$\mu$m will give a combined series resistance of above 60\,k$\Omega$ at 50\,mK. The films we deposited were found to be amorphous, with no magnetic phases present, which means they are suitable for use in superconducting circuits. 

For fabrication of an on-chip circuit, gold or a similarly low-resistivity intermediate layer should be used to ensure a low contact resistance. Rapid oxidation of the surface of the films means careful consideration of the method of device manufacture and order of the layers is also needed.

For minimising temperature rises in a circuit at mK temperatures, strongly oxidised chromium-oxide resistors can be favourable compared with weakly oxidised resistors with the same value of resistance \cite{Webster}. We also should verify that the parasitic capacitance does not result in a cut-off frequency too low for QPS applications: Using a value of capacitance per unit area for the resistors determined by Zorin et al. \cite{Zorin}, we find a cut-off frequency of 5\,GHz. This may be sufficiently high that the circuit dynamics are not affected.

We conclude that chromium oxide films are suitable for use for thin-film resistors with values in the range of hundreds of ohms to mega-ohms, and in particular are suited to use in applications such as QPS circuits.

\section{Acknowledgments}

The work is supported in part by EPSRC (Grant No. EP/H0055441 and Grant No. EP/J017329/1) and by Raith GmbH. The authors wish to thank Kevin Reeves from UCL Institute of Archaeology for his assitance with WDS measurements.


\begin{thebibliography}{99}
\bibitem{Footnote} It is worth noting that some conduction through the semiconducting substrate is likely in such measurements.
\end{thebibliography}
\end{document}